\documentstyle[11pt]{article}

\catcode`\@=11

\def\footnotesize{\@setsize\footnotesize{11pt}\ixpt\@ixpt
       \abovedisplayskip \z@
      \belowdisplayskip\z@
     \abovedisplayshortskip\abovedisplayskip
    \belowdisplayshortskip\belowdisplayshortskip
\def\@listi{\leftmargin\leftmargini \topsep 3pt plus 1pt minus 1pt
     \parsep 2pt plus 1pt minus 1pt
    \itemsep \parsep}}

\skip\footins 12pt plus 12pt

\def\footnoterule{\kern3\p@  \hrule width 3em\vspace{3pt}} 

\def\ps@plain{\let\@mkboth\@gobbletwo
     \def\@oddfoot{{\hfil\small\thepage\hfil}}%
     \def\@oddhead{}
      \def\@evenhead{}\def\@evenfoot{}}

\def\ps@headings{\let\@mkboth\markboth
        \def\@oddfoot{}\def\@evenfoot{}%
        \def\@evenhead{{\rm\thepage}\hspace*{2pc}{\sc\leftmark}\hfil}%
        \def\@oddhead{\hfil{\noindent\sc\rightmark}\hspace*{2pc}{\rm\thepage}}%

\def\ps@myheadings{\let\@mkboth\@gobbletwo
 \def\@oddfoot{}\def\@evenfoot{}%
 \def\@oddhead{\hfil{\sc\rightmark}\hspace*{2pc}{\normalsize\rm\thepage}}%
 \def\@evenhead{{\normalsize\rm\thepage}\hspace*{2pc}{\sc\leftmark}\hfil}%
}}

\def\abstract{\if@twocolumn
\section*{Abstract}
\else \small
\begin{center}
{\bf Abstract\vspace{-.5em}\vspace{3pt}}
\end{center}
\quotation
\fi}
\def\endabstract{\if@twocolumn\else\endquotation\fi}

\def\newproof#1{\@nprf{#1}}

\def\@nprf#1#2{\@xnprf{#1}{#2}}

\def\@xnprf#1#2{\expandafter\@ifdefinable\csname #1\endcsname
\global\@namedef{#1}{\@prf{#1}{#2}}\global\@namedef{end#1}{\@endproof}}

\def\@prf#1#2{\@xprf{#1}{#2}}

\def\@xprf#1#2{\@beginproof{#2}{\csname the#1\endcsname}\ignorespaces}

\def\newalgorithm#1{\@ifnextchar[{\@oalg{#1}}{\@nalg{#1}}}

\def\@nalg#1#2{%
\@ifnextchar[{\@xnalg{#1}{#2}}{\@ynalg{#1}{#2}}}

\def\@xnalg#1#2[#3]{\expandafter\@ifdefinable\csname #1\endcsname
{\@definecounter{#1}\@addtoreset{#1}{#3}%
\expandafter\xdef\csname the#1\endcsname{\expandafter\noexpand
  \csname the#3\endcsname \@thmcountersep \@thmcounter{#1}}%
\global\@namedef{#1}{\@alg{#1}{#2}}\global\@namedef{end#1}{\@endalgorithm}}}

\def\@ynalg#1#2{\expandafter\@ifdefinable\csname #1\endcsname
{\@definecounter{#1}%
\expandafter\xdef\csname the#1\endcsname{\@thmcounter{#1}}%
\global\@namedef{#1}{\@alg{#1}{#2}}\global\@namedef{end#1}{\@endalgorithm}}}

\def\@oalg#1[#2]#3{\expandafter\@ifdefinable\csname #1\endcsname
  {\global\@namedef{the#1}{\@nameuse{the#2}}%
\global\@namedef{#1}{\@alg{#2}{#3}}%
\global\@namedef{end#1}{\@endalgorithm}}}

\def\@alg#1#2{\refstepcounter
    {#1}\@ifnextchar[{\@yalg{#1}{#2}}{\@xalg{#1}{#2}}}

\def\@xalg#1#2{\@beginalgorithm{#2}{\csname the#1\endcsname}\ignorespaces}
\def\@yalg#1#2[#3]{\@opargbeginalgorithm{#2}{\csname
       the#1\endcsname}{#3}\ignorespaces}

\def\@beginproof#1{\rm {\it #1.\ }}
\def\@endproof{\outerparskip 0pt\endtrivlist}

\def\@begintheorem#1#2{\it {\sc #1\ #2.\ }}
\def\@opargbegintheorem#1#2#3{\it
      {\sc #1\ #2\ (#3).\ }}
\def\@endtheorem{\outerparskip 0pt\endtrivlist}

\def\@beginalgorithm#1#2{\rm \trivlist \item[\hskip \labelsep{\sc #1\ #2.}]}
\def\@opargbeginalgorithm#1#2#3{\rm \trivlist
      \item[\hskip \labelsep{\sc #1\ #2.\ (#3)}]}
\def\@endalgorithm{\outerparskip 6pt\endtrivlist}

\newskip\outerparskip

\def\trivlist{\parsep\outerparskip
  \@trivlist \labelwidth\z@ \leftmargin\z@
  \itemindent\parindent \def\makelabel##1{##1}}

\def\@trivlist{\topsep=0pt\@topsepadd\topsep
  \if@noskipsec \leavevmode \fi
  \ifvmode \advance\@topsepadd\partopsep \else \unskip\par\fi
  \if@inlabel \@noparitemtrue \@noparlisttrue
    \else \@noparlistfalse \@topsep\@topsepadd \fi
    \advance\@topsep \parskip
  \leftskip\z@\rightskip\@rightskip \parfillskip\@flushglue
  \@setpar{\if@newlist\else{\@@par}\fi}%
  \global\@newlisttrue \@outerparskip\parskip}

\def\endtrivlist{\if@newlist\@noitemerr\fi
   \if@inlabel\indent\fi
   \ifhmode\unskip \par\fi
   \if@noparlist \else
      \ifdim\lastskip >\z@ \@tempskipa\lastskip \vskip -\lastskip
         \advance\@tempskipa\parskip \advance\@tempskipa -\@outerparskip
         \vskip\@tempskipa
   \fi\@endparenv\fi
   \vskip\outerparskip}

 \newproof{@proof}{Proof}

 \newtheorem{@theorem}{Theorem}[section]

\newproof{Example}{Example}
\newproof{Method}{Method}
\newproof{Exercise}{Exercise}

 \def\@figtxt{figure}
\long\def\@makecaption#1#2{\small
\setlength{\parindent}{18pt}
\baselineskip 14pt
 \ifx\@captype\@figtxt
 \vskip 10pt
 \setbox\@tempboxa\hbox{{\sc #1} {\it #2}}
 \ifdim \wd\@tempboxa >\hsize {\sc #1} {\it #2}\par \else \hbox
to\hsize{\hfil\box\@tempboxa\hfil}%
 \fi\else\hbox to\hsize{\hfil{\sc #1}\hfil}%
 \setbox\@tempboxa\hbox{{\it #2}}%
 \ifdim \wd\@tempboxa >\hsize {\it #2}\par \else
 \hbox to \hsize{\hfil\box\@tempboxa\hfil}\fi
 \vskip 10pt
 \fi}

\def\fnum@figure{\par\sc Fig. \thefigure.\ }
\def\fnum@table{\small \sc Table \thetable}

\def\section{\@startsection {section}{1}{\z@}{-3.5ex plus -1ex minus
 -.2ex}
{2pt}{\large\bf}}
\def\subsection{\@startsection{subsection}{2}{\z@}{-3.25ex plus -1ex minus
 -.2ex}
{2pt}{\large\bf}}
\def\subsubsection{\@startsection {subsubsection}{3}{\z@}{1.3ex plus .5ex minus
    .2ex}{-.5em plus -.1em}{\normalsize\bf}}

\def\thebibliography#1{%
\parindent 0em
\vspace{9pt}
\begin{flushleft}\large\bf {References}\end{flushleft}
\addvspace{3pt}\nopagebreak\list
{[\arabic{enumi}]} {\settowidth\labelwidth{[#1]}
\leftmargin\labelwidth
\leftmargin=17pt
 \advance\leftmargin\labelsep
 \usecounter{enumi}\@bibsetup}
\def\newblock{\hskip .11em plus .33em minus -.07em}
 \sloppy\clubpenalty4000\widowpenalty4000
 \sfcode`\.=1000\relax}

\def\@bibsetup{
\itemsep=0pt \parsep=0pt
\small}

\def\theindex{\@restonecoltrue\if@twocolumn\@restonecolfalse\fi
\columnseprule \z@
\columnsep 35pt\twocolumn[\chapter*{Index}]
 \parskip\z@ plus .3pt\relax\let\item\@idxitem}

\def\printindex{\cleardoublepage\markboth{INDEX}{INDEX}
\addcontentsline{toc}{chapter}{Index}\@input{\jobname.ind}}

\ps@headings

\def\Tr{\mathop{\rm Tr}}

\def\so{{\hbox{${\rm SO}(2,1)$}}}

\def\iso{{\hbox{${\rm ISO}(2,1)$}}}
\def\isoc{{\hbox{${\rm ISO}_0(2,1)$}}}

\def\isot{{\hbox{${\rm ISO}(3)$}}}

\def\miso{{{\cal M}}}
\def\ms{{{\cal M}_s}}
\def\mt{{{\cal M}_t}}
\def\mn{{{\cal M}_n}}
\def\mo{{{\cal M}_0}}

\begin{document}
\cleardoublepage
\pagestyle{myheadings}

\title{Global structure of Witten's 2+1 gravity on ${\bf R}\times T^2$}
\author{Jorma Louko\thanks{Department of Physics,
University of Wisconsin--Milwaukee,
P.O.\ Box 413,
Milwaukee, Wisconsin 53201.
Supported by NSF Grants PHY90-16733 and PHY91-05935.}
\and
Donald Marolf\thanks{Center for Gravitational Physics and Geometry,
Physics Department,
Penn State University, University Park, PA 16802-6300.
Supported by NSF Grants PHY90-05790 and PHY93-96246.}
}
\date{February 1994\thanks{Contribution to the proceedings of the
Cornelius Lanczos International Centenary Conference,
Raleigh, NC, December 12--17, 1993.}}
\maketitle
\markboth{Louko and Marolf}{Witten's 2+1 Gravity on the Torus}

\vglue-5 truecm
\vbox{\baselineskip=12pt
\rightline{CGPG-94/3-2}
\rightline{WISC-MILW-94-TH-12}
\rightline{gr-qc/9403041}
}
\vglue4 truecm

\pagenumbering{arabic}

\begin{abstract}
We investigate the space $\miso$ of classical solutions to Witten's
formulation of 2+1 gravity on the manifold ${\bf R} \times T^2$. $\miso$ is
connected, but neither Hausdorff nor a manifold. However, removing from
$\miso$ a set of measure zero yields a connected manifold which is naturally
viewed as the cotangent bundle over a non-Hausdorff base space. Avenues
towards quantizing the theory are discussed in view of the relation between
spacetime metrics and the various parts of~$\miso$.
\end{abstract}

\section{Introduction}

The observation that vacuum Einstein gravity in 2+1 spacetime dimensions
has no local dynamical degrees of freedom has stimulated
interest in 2+1 gravity as an arena where quantum gravity can be
investigated without many of the technical complications that are
present in 3+1 spacetime dimensions. Of particular interest for the 3+1
theory is the relationship between the various 2+1 quantum
theories that have been constructed in the metric, connection, and loop
formulations. For a recent review, see~\cite{carlip-water}.

In this contribution we shall consider Witten's formulation of 2+1 gravity
\cite{achu,witten1} and its relation to the conventional metric formulation.
On manifolds of the form ${\bf R}\times \Sigma$, where $\Sigma$ is a closed
orientable surface of genus~$g>1$, the situation is well understood: the
space of classical solutions to Witten's theory contains several
disconnected components, one of which is a smooth manifold isomorphic to the
solution space of the conventional metric formulation
\cite{witten1,moncrief,mess}. On the manifold ${\bf R}\times S^2$ the
situation is trivial, in the sense that Witten's theory possesses only one
classical solution and the conventional metric formulation possesses no
solutions \cite{moncrief,mess}. Our aim is to describe the solution space to
Witten's theory on the manifold ${\bf R}\times T^2$, and to explore the
avenues that the global structure of this solution space offers for
quantizing the theory. The details and more references to earlier work can
be found in \cite{louma}.

\section{Outline of results}

Recall that Witten's formulation \cite{achu,witten1} of 2+1
gravity on an orientable manifold $M$ can be derived from the action
$S \left( {\bar e} , {\bar A} \right)
= \int_{M} \Tr \left({\bar e} \wedge {\bar F} \right)$.
Here ${\bar e}$ is a co-triad taking values in the dual of the Lie
algebra of $\so$, ${\bar F}$ is the curvature of the $\so$
connection~${\bar A}$, and the trace refers to a contraction
in the Lie algebra indices. For $M={\bf R}\times \Sigma$, where $\Sigma$ is
a closed orientable surface, the pull-backs of ${\bar e}$ and ${\bar A}$ to
$\Sigma$ define an $\iso$ connection ${\cal A}$ on~$\Sigma$. Witten
\cite{witten1} observed that the equations of motion enforce ${\cal A}$ to
be flat, and further that the dynamics consists entirely of gauge
transformations of~${\cal A}$. Choosing ${\cal A}$ to live on the trivial
principal bundle $\isoc\times\Sigma$, where $\isoc$ is the connected
component of $\iso$, one thus sees that the space of classical solutions is
just the moduli space of flat $\isoc$ connections on~$\Sigma$, modulo
$\isoc$ gauge transformations. This space can be described as the
space of group homomorphisms from the fundamental group of $\Sigma$ to
$\isoc$, modulo overall conjugation by $\isoc$ \cite{carlip-water}.

We now focus on the case where $\Sigma$ is the torus~$T^2$. Let $\miso$
denote the space of classical solutions to Witten's theory. As the
fundamental group of the torus is the abelian group ${\bf Z}\times {\bf Z}$,
we see from the above that the points in $\miso$ are just equivalence
classes of pairs of commuting elements of $\isoc$ modulo $\isoc$
conjugation. We need to give a characterization of such equivalence classes.

Suppose for the moment that we were considering Euclidean rather than
Lorentzian gravity. In this case $\isoc$ would be replaced by $\isot$, which
is the group of rigid body motions in three dimensional Euclidean space.
Now, a classic result known as Euler's theorem says that an element of
$\isot$ can always be written as a rotation about some axis followed by a
translation along the same axis. If two elements of $\isot$, not both purely
translational, are written in this fashion and then required to commute, one
finds that the  respective axes of rotation must be the same axis, and by
conjugation this axis can be chosen to be (say) the $z$-axis. One thus
recovers a space whose points are parametrized by the two rotation angles
and the magnitudes of the two translations, modulo certain identifications
which stem from further conjugation by a rotation by $\pi$ about the
$x$-axis. The special case where both $\isot$ elements are purely
translational requires a separate consideration; such points are
parametrized by the magnitudes of the two translations and the angle between
them, again modulo certain identifications. The space is thus
roughly speaking four dimensional, but not quite a manifold.

Return now to Lorentzian gravity. The crucial difference between $\isot$ and
$\isoc$ for us is, of course, that there are (apart from the identity
rotation) three distinct types of Lorentz rotations in $\isoc$: the boosts,
which fix a spacelike axis, the rotations, which fix a timelike axis, and
the null rotations, which fix a null axis. For rotations and boosts there
are natural analogues of Euler's theorem, and the analysis proceeds fairly
similarly to that in the Euclidean case. The rotational part $\mt$ of
$\miso$ can be understood as the cotangent bundle over the punctured torus:
the base space arises from the Lorentz components and the cotangent fibers
arise from the translational components of the $\isoc$ elements. The
symplectic structure that allows the interpretation of this space as a
cotangent bundle arises from the Hamiltonian decomposition of the action.
The boost part $\ms$ of $\miso$ can be similarly understood as the cotangent
bundle over the punctured plane. For null rotations, however, there is no
direct analogue of Euler's theorem. The null part $\mn$ of $\miso$ turns out
to be a three dimensional manifold with topology $S^1\times {\bf R}^2$, the
factor $S^1$ coming from the Lorentz components and the factor ${\bf R}^2$
coming from the translational components of the $\isoc$ elements.  The
remaining part $\mo$ of $\miso$ consists of the case where both $\isoc$
elements are purely translational. $\mo$ is close to being a  three
dimensional manifold, but its structure is complicated by the different
possibilities for the spacelike/timelike/null character of the plane or line
(or point) which the two translation vectors span.

A first observation is that $\miso$ is a {\em connected\/} space. This is
in a striking contrast with the situation in Witten's theory on manifolds
where the torus is replace by a higher genus surface \cite{witten1}. Indeed,
one can view $\miso$ as the two four dimensional manifolds $\mt$ and $\ms$
glued together by the three dimensional manifold $\mn$ and the set~$\mo$.
The gluing is not smooth, and $\miso$ itself is neither Hausdorff nor a
manifold. However, the connected set $\miso\setminus\mo=\mt\cup\mn\cup\ms$,
which contains all of $\miso$ except a set of measure zero, {\em is\/} a
manifold: it can be viewed as the cotangent bundle whose base space consists
of the base space of~$\ms$ (punctured plane) and the base space of~$\mt$
(punctured torus) glued together at the punctures; the circle which provides
the glue is the $S^1$ factor of~$\mn$. The circle joins to the base space of
$\mt$ in a one-to-one fashion, but the joining of the base space of $\ms$ to
the circle is two-to-one. This makes $\miso\setminus\mo$ a non-Hausdorff
manifold.

The above structure of $\miso$ suggests various avenues for quantizing the
theory. One possibility is to quantize $\ms$ and $\mt$ separately; this
leads to the theories considered in \cite{five-a,AAbook2,carlip1}.
However, it is also possible to perform a quantization on all of
$\miso\setminus\mo$ at once. The resulting larger theory contains the
theories of \cite{five-a,AAbook2,carlip1} as its parts,
and in particular it contains operators that induce transitions between
these smaller theories.

\section{Discussion}

Is there any physical interest in the large quantum theory constructed on
all of $\miso\setminus\mo$?  We would like to end by speculating that this
question may be related to the role of closed timelike loops in quantum
gravity.

Recall that in the conventional metric formulation of 2+1 gravity one
assumes that the spacetime metric is nondegenerate, and in the Hamiltonian
decomposition on the manifold ${\bf R}\times T^2$ one further assumes that
the induced metric on $T^2$ is spacelike \cite{moncrief}. When such
spacetimes are mapped to Witten's description, the image ${\cal M}_{\rm
metric}$ lies in $\ms\cup\mo$, filling most of $\ms$ but only roughly half
of~$\mo$ \cite{moncrief,mess}.  However, it can be shown \cite{louma} that
for any point in~$\miso$, with the exception of a set of measure zero, there
exist corresponding nondegenerate spacetime metrics on ${\bf R}\times T^2$.
For points in $\miso$ that are not in ${\cal M}_{\rm metric}$, such a
nondegenerate spacetime will necessarily contain closed timelike or null
loops. This suggests viewing the large quantum theory described above as a
theory containing, in some rough sense, transitions between spacetimes with
closed causal loops and spacetimes without closed causal loops. It would be
of interest to understand whether this speculation could be augmented into a
more concrete statement.

\end{document}